\documentclass[graybox]{svmult}

\usepackage{mathptmx}       
\usepackage{helvet}         
\usepackage{courier}        
\usepackage{type1cm}        
\usepackage{makeidx}         
\usepackage{graphicx}        
\usepackage{multicol}        
\usepackage[bottom]{footmisc}
\usepackage[latin9]{inputenc}

\makeindex             

\begin{document}

\title*{ Shot Noise in Digital Holography }
\author{Fadwa~Joud \and
        Frédéric~Verpillat \and
        Michael~Atlan \and
        Pierre-André~Taillard \and
        Michel~Gross}
\institute{
Fadwa~Joud \at  Laboratoire Kastler Brossel \'Ecole Normale Sup\'erieure , UMR 8552 , UPMC, CNRS 24 rue
Lhomond , 75231 Paris Cedex 05. \email{joud@lkb.ens.fr}
\and Frédéric~Verpillat  \at Laboratoire Kastler Brossel \'Ecole Normale Sup\'erieure , UMR 8552 , UPMC, CNRS 24 rue
Lhomond , 75231 Paris Cedex 05, \email{verpillat@lkb.ens.fr}
\and Michael~Atlan \at Fondation Pierre-Gilles de Gennes, Institut Langevin: UMR 7587 CNRS,  U 979 INSERM, ESPCI ParisTech, Universit\'e Paris 6, Universit\'e Paris 7, 10 rue Vauquelin, 75 231 Paris Cedex 05, France.\email{atlan@optique.espci.fr}
\and Pierre-André~Taillard \at Conservatoire de musique neuchâtelois; Avenue Léopold-Robert 34 ; 2300 La Chaux-de-Fonds
; Suisse.\email{taillard@hispeed.ch}
 \and Michel~Gross \at  Laboratoire Kastler Brossel \'Ecole
Normale Sup\'erieure , UMR 8552 , UPMC, CNRS 24 rue Lhomond , 75231 Paris Cedex 05. \email{gross@lkb.ens.fr}
 }
%
%
\maketitle

\abstract{We  discuss  on  noise in heterodyne holography in an off-axis configuration. We show that, for a weak signal, the noise is dominated by the shot noise on the reference beam. This noise corresponds to an equivalent noise on the signal beam  of 1 photo electron per pixel,
for the whole sequence of images used to build the digital hologram.}

\section{Introduction}
\label{sec:1}
%
%

Demonstrated by Gabor \cite{Gabor49} in the early 50's, the purpose of holography is to record, on a 2D detector, the
phase and the amplitude of the radiation field scattered by an object under coherent illumination. The photographic film used in conventional holography is replaced by a 2D electronic detection in digital holography \cite{Macovsky1971} enabling quantitative numerical analysis. Digital holography has been waiting for
the recent development of computer and video technology to be experimentally demonstrated \cite{Schnars94}. The main
advantage of digital holography is that, contrary to holography with photographic plates \cite{Gabor49}, the holograms
are recorded by a CCD, and the image is digitally reconstructed by a computer, avoiding photographic processing
\cite{Goodmann_1967}.

Off-axis holography \cite{leith1965microscopy} is the oldest configuration adapted to digital
holography \cite{Schnars_Juptner_94,Schnars94,Kreis88}. In off-axis digital holography, as well as in photographic
plate holography, the reference or local oscillator (LO) beam is angularly tilted with respect to the object
observation axis. It is then possible to record, with a single hologram, the two quadratures of the object's complex
field. However, the object field of view is reduced, since one must avoid the overlapping of the image with the
conjugate image alias \cite{Cuche00}. In Phase-shifting digital holography, which has been introduced later
\cite{Yamaguchi1997}, several images are recorded with different LO beam phases. It is then possible to obtain the
two quadratures of the object field in an in-line configuration even though the conjugate image alias and the true image
overlap, because aliases can be removed by taking image differences.

%
%

We have developed an alternative phase-shifting  digital holography
technique, called heterodyne holography,  that uses a frequency
shift of the reference beam to continuously shift the phase of the
recorded interference pattern \cite{Leclerc2000}. One of the
advantages of this technique is its ability to provide accurate
phase shifts that allow to suppress twin images aliases
\cite{atlan2007accurate}. This greatly simplifies holographic data
handling, and improves sensitivity. Moreover, it is possible to
perform holographic detection at a frequency different from
illumination. One can for example detect "tagged photons"
\cite{gross2003shot,atlan2005pulsed} in ultrasound-modulated optical
imaging \cite{wang1997ultrasound}.   One can also also image
vibrating objects at the frequencies corresponding to vibration
sidebands \cite{joud2009imaging,joud2009fringe}. To the end, it is
possible to  perform Laser Doppler imaging \cite{atlan2006laser}
within microvessels \cite{atlan2006frequency,atlan2007cortical,atlan2008high}.

More generally, our setup can be viewed as a multipixel heterodyne
detector that is able of recording the complex amplitude of the
signal electromagnetic field $\cal E$ onto all pixels of the CCD
camera in parallel. We get the map of the field over the array detector (i.e. ${\cal E}(x,y)$ where $x$ and $y$ are the pixels
coordinates). Since the field is measured on all pixels at the same
time, the relative phase that is measured for different locations
$(x,y)$ is meaningful. This means that the field map ${\cal E}(x,y)$
is a hologram that can be used to reconstruct the field $\cal E$ in
any location, in particular in the object plane.

%

In the present paper we will discuss on  noise in digital
holography, and we will try to determine what is the ultimate noise
limit both theoretically, and in real time holographic experiments.
We will see that, in the theoretical ideal case, the limiting noise
is the Shot Noise on the holographic reference beam. In reference to heterodyne detection, the reference beam is also called Local Oscillator. We will see that the ultimate theoretical limiting noise
can be reached in real time holographic experiment, by using
heterodyne holography \cite{Leclerc2000} in off-axis configuration.
This combination makes possible to fully filter off the technical
noise, whose main origin is the LO beam technical noise, opening the
way to holography with ultimate sensitivity
\cite{gross2007dhu,gross2008noise}.

\section{Theoretical noise}\label{section_Theoretical_noise}

To discuss on noise in  digital digital holography, we will consider
both  the case of off-axis holography, where the hologram is
obtained from one frame of the CCD camera, and the case of  phase
shifting holography, where the holographic information is extracted
from a sequence of $M$ frames.

We will thus consider a sequence of $M$ frames: $I_0$ to $I_{M-1}$
(where $M=1$ in the one shot, off axis case). For each frame $I_k$, let us
note $I_{k,p,q}$ the  CCD camera signal on each pixel, where $k$ is
the frame index, and $p,q$ the pixel indexes along the $x$ and $y $
directions. The CCD signal  $I_{k,p,q}$ is measured in Digital Counts
(DC) units. In the typical case of the 12 bit digital camera used in
experiments below, we have  $0 \leq I_{k,p,q}<4096$. For each frame
$k$, the optical signal is integrated by  over the acquisition time
$T=1/f_{\rm{ccd}}$ of the CCD camera. The pixel signal $I_{k,p,q}$
is thus defined by :
\begin{equation}\label{Eq_slide_1}
I_{k,p,q}=\int_{t_k-T/2}^{t_k+T/2} {dt}\; \int\!\int_{(p,q)} {dx}{dy} \;\;|E(x,y,t) +E_{LO}(x,y,t)|^2
\end{equation}
where  $\int\!\int_{(p,q)} dx dy $ represents  the integral over the
pixel $(p,q)$ area, and where $t_k$ is  the recording moment of
frame $k$. Introducing the complex representations ${\cal E}$ and
${\cal E}_{LO}$ of the fields $E$ and $E_{LO}$, we get :
\begin{equation}
E(x,y,t)={\cal E}(x,y)e^{j\omega_I t}+c.c.
\end{equation}
\begin{equation}
E_{LO}(x,y,t)={\cal E}_{LO}(x,y)e^{j\omega_{LO} t}+c.c
\end{equation}
\begin{equation}\label{Eq_slide_2}
I_{k,p,q}=a^2T \left(|{\cal E}_{p,q}|^2+|{\cal E}_{LO}|^2+{\cal E}_{p,q} {\cal E}_{LO}^*\cdot e^{ \left(j
\left(\omega_I-\omega_{LO}\right)t_k\right)}+c.c.\right)
\end{equation}
where $a$ is the pixel size. To simplify the notations in
Eq.\ref{Eq_slide_2}, we have considered that the LO field ${\cal
E}_{LO}$ is the same in all locations $(x,y)$, and that signal
field ${\cal E}_{p,q}$  does not vary within the pixel $(p,q)$.
If ${\cal E}_{LO}$ varies with location, one has to replace
${\cal E}_{LO}$ by ${\cal E}_{LO,p,q}$ in Eq.\ref{Eq_slide_2}.

In the single-shot, off-axis holography case, the hologram $H$ is simply $H
\equiv I_k$. In order to simplify the discussion in the phase
shifting digital holography case \cite{Yamaguchi1997}, we will
consider 4 phases holographic detection ($M=4n$). In that case, the
phase shift of the LO beam  equal to $\pi/2$ from one recorded frame to the
next. Because of this shift, the complex hologram $H$ is obtained by
summing the sequence of $M$ frames $I_1$ to $I_{M}$ with the
appropriate phase coefficient :

\begin{equation}\label{Sum_heterodyne}
H  \equiv \sum^{M}_{k=1} (j)^{k-1} I_k
\end{equation}
where $H$ is a matrix of pixel $H_{p,q}$, and where  $M=4n$ in the
4-phases phase-shifting case, and $M=1$ in the single-shot, off-axis case. We get from Eq.\ref{Eq_slide_2} :
\begin{equation}\label{Sum_heterodyne_2}
H_{p,q}= \sum^{M}_{k=1} (j)^{k} I_{k,p,q} = 4n a^2 T {\cal E}_{p,q} {\cal E}_{LO}^*
\end{equation}
The complex hologram $H_{p,q}$ is thus proportional to the
object field ${\cal E}_{p,q}$ with a proportionality factor that
involves  ${\cal E}_{LO}^*$.

\subsection{The Shot Noise on the CCD pixel signal}

Because of spontaneous emission, laser emission and photodetection are random processes, the signal that is obtained on a CCD pixel exhibits a Poisson noise called ''shot noise''. The
effect of this Poisson noise on the signal, and on the
holographic images, is the Ultimate Theoretical
Limiting noise, which we will study here.

We can split the  signal  $I_{k,p,q}$:  we get for  frame $k$ and
pixel $(p,q)$, in a noiseless  average component $\langle I_{k,p,q}
\rangle $  (here $\langle ~\rangle$ is the statistical average
operator) and a noise component $i_{k,p,q}$:
\begin{equation}\label{EQ_noise on Ikpq}
    I_{k,p,q} \equiv  \langle I_{k,p,q} \rangle + i_{k,p,q}
\end{equation}
To go further in the discussion, we will use photo electrons Units
to measure the signal $I_{k,p,q}$.

We must notice that the local oscillator signal ${\cal{E}}_{LO}$ is
large, and corresponds to a large number of photo electrons (e). In
real life, this assumption is true. For example, if we adjust the
power of the LO beam to be at the half maximum for the camera signal
in DC unit (2048 DC for a 12 bits camera), the pixel signal will be
about $10^4$~e for the camera used below in experiments, since  the
''Camera Gain'' is $4.8~$e per DC. This yields two consequences,
which simplify the analysis.
First, the signal $I_{k,p,q} $ exhibits a gaussian distribution around its statistical average.
Second, both the quantization noise of  the photo electron signal ($I_{k,p,q}$ is an integer in photo electron Units),
and the quantization noise of   the Digital Count signal ($I_{k,p,q}$ is an integer in DC Units) can be neglected.
These approximations are valid, since the width of the $I_{k,p,q} $ gaussian distribution is much larger than one in
both photo electron and DC Units. In the example given above, $\langle I_{k,p,q} \rangle \simeq 10^4$, and this width
is $\simeq 10^2$ in photo electron Units, and $\simeq 20$ in DC Units.
%
%
One can thus consider that  $I_{k,p,q} $, $\langle I_{k,p,q}
\rangle$ and $i_{k,p,q}$ are  floating numbers (and not integer).
Moreover, $i_{k,p,q}$ is a random Gaussian distribution, with :
\begin{equation}\label{Eq_Ikpq variance_0}
   \langle i_{k,p,q} \rangle = 0
\end{equation}
\begin{equation}\label{Eq_Ikpq variance}
   \langle i_{k,p,q}^2 \rangle = \langle I_{k,p,q} \rangle
\end{equation}

To analyse the shot noise's contribution to the holographic signal
$H_{p,q}$, one of the most simple method is to perform Monte Carlo
simulation from Eq.\ref{EQ_noise on Ikpq}, Eq.\ref{Eq_Ikpq
variance_0} and Eq.\ref{Eq_Ikpq variance}. Since $I_{k,p,q}$ is ever
large in real life (about $10^4$ in our experiment),  $\langle
I_{k,p,q} \rangle $ can be replaced by $ I_{k,p,q} $ (which is
measured in experiment)   in the right member of Eq.\ref{Eq_Ikpq
variance}. One has thus:
\begin{equation}\label{Eq_Ikpq variance2}
   \langle i_{k,p,q}^2\rangle = \langle I_{k,p,q} \rangle \simeq I_{k,p,q}
\end{equation}
Monte Carlo simulation of the noise can be done from Eq.\ref{EQ_noise
on Ikpq}, Eq.\ref{Eq_Ikpq variance_0}  and Eq.\ref{Eq_Ikpq
variance2}


\subsection{The Object field Equivalent Noise  for 1 frame}

\begin{figure}[]
\sidecaption
\includegraphics[width=7cm,keepaspectratio=true]{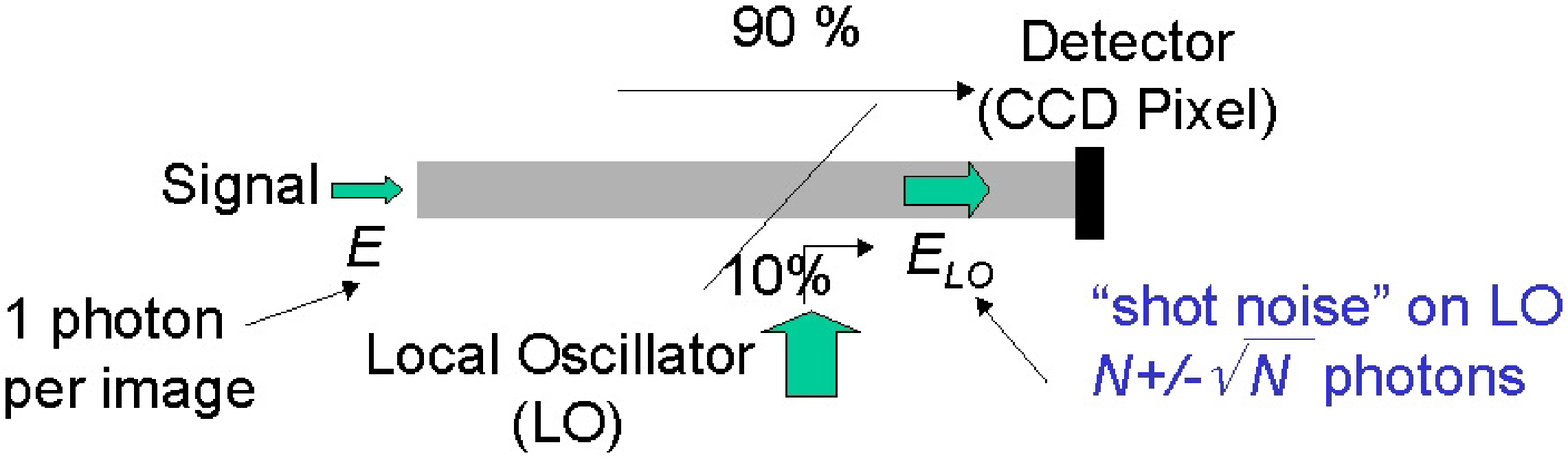}
\caption{ 1 photon equivalent signal (accounting Heterodyne gain),  and shot noise on the holographic Local Oscillator
beam. }\label{fig2_shot_noise}
\end{figure}

In order to discuss the effect of the shot noise on the heterodyne signal ${\cal E}_{p,q} {\cal E}_{LO}^*$ of
Eq.\ref{Eq_slide_2}, let us consider the simple situation sketched on Fig.\ref{fig2_shot_noise}. A weak object field $
{{E}}$, with  1 photon or 1 photo electron per pixel and per frame, interferes with a LO field $ {{E}}_{LO}$ with $N$
photons, where $N$ is large ($N = 10^4$, in the case of our experiment). Since the LO beam signal $a^2T|{\cal
E}_{LO}|^2$ is equal to $N $ photons, and the object field signal  $a^2T|{\cal E}_{p,q}|^2$ is one photon, we have:
\begin{equation}\label{Eq_equiv_noise_1}
    I_{k,p,q} = N + 1 + i_{k,p,q} + a^2T{\cal E}_{p,q} {\cal E}_{LO}^* e^{...} + c.c.
\end{equation}
The heterodyne signal ${\cal E}_{p,q} {\cal E}_{LO}^*$ is much
larger than $|{\cal E}_{p,q}|^2$. This is the gain effect,
associated to the  coherent detection of the field ${\cal E}_{p,q}$.
This gain is commonly called ''heterodyne gain'', and is
proportional to the amplitude of the LO field ${\cal E}_{LO}^*$.

The purpose of the present discussion is to determine the effect of
the noise term $i_{k,p,q}$ in  Eq.\ref{Eq_equiv_noise_1} on the
holographic signal $H_{p,q}$. Since $H_{p,q}$ involves only the
heterodyne term ${\cal E}_{p,q} {\cal E}_{LO}^*$ (see
Eq.\ref{Sum_heterodyne_2}), we have to compare, in
Eq.\ref{Eq_equiv_noise_1},
   the shot noise term $i_{k,p,q}$,
   and the heterodyne term ${\cal E}_{p,q} {\cal E}_{LO}^*$.
   %

Consider first the shot noise term.   We have
\begin{equation}\label{Eq_equiv_noise_3}
    \langle i_{k,p,q}^2 \rangle  = \langle I_{k,p,q}\rangle = N + 1 \simeq N
\end{equation}
The variance of the shot noise term is thus $N^{1/2}$. Since this
noise is mainly related to the shot noise on the local oscillator
(since $N\gg 1$), one can group together, in
Eq.\ref{Eq_equiv_noise_1}, the LO beam term (i.e. $N$) with the
noise term $i_{k,p,q}$, and consider that the LO beam signal
fluctuates,  the number of LO beam photons being thus ''$N \pm
N^{1/2}$'', as mentioned on Fig.\ref{fig2_shot_noise}.

Consider now the the heterodyne beat signal. Since we have $N$ photons on the LO beam, and 1 photon on the object beam,
we get:
\begin{equation}\label{Eq_equiv_noise_2}
    a^2T |{\cal E}_{p,q} {\cal E}_{LO}^*| \equiv \left( \left(a^2T{|\cal E}_{p,q}|^2\right) \left(a^2T{|\cal
    E}_{LO}|^2\right)\right)^{1/2}=N^{1/2}
\end{equation}
The heterodyne beat signal ${\cal E}_{p,q} {\cal E}_{LO}^*$  is thus
$N^{1/2}=100$.

The shot noise term  $ i_{k,p,q}$  is thus equal to the heterodyne
signal ${\cal E}_{p,q} {\cal E}_{LO}^*$ corresponding to 1 photon on
the object field. This means that shot noise $ i_{k,p,q}$ yields  an
equivalent noise of 1 photon per pixel, on the object beam. This
result is obtained here for 1 frame. We will show that it remains
true for a sequence of $M$ frames, whatever $M=4n$ is.

\subsection{The Object field Equivalent Noise  for  $M=4n$ frames}

Let us introduce the  DC component signal $D$, which is similar to the heterodyne signal $H$ given by
Eq.\ref{Sum_heterodyne}, but without phase factors:
\begin{equation}\label{DC_term}
D \equiv \sum_{k=1}^{M} I_k
\end{equation}
The component $D$ can be defined for each pixel $(p,q)$ by :
\begin{equation}\label{DC_term_pq}
D_{p,q} \equiv \sum_{k=1}^{M} I_{k,p,q}
\end{equation}
%
Since $I_{k,p,q}$ is always large in real life (about $10^4$ in our experiment), the shot noise term can be neglected  in
the calculation of $ D_{p,q}$ by  Eq.\ref{DC_term_pq}. We have thus:
\begin{equation}\label{DC_term_pq_1}
D_{p,q} \equiv  \sum_{k=1}^{M} I_{k,p,q}   = M a^2T \left(|{\cal E}_{p,q}|^2+|{\cal E}_{LO}|^2\right)
\end{equation}
We are implicitly interested  by the
low signal situation (i.e. ${\cal E}_{p,q}\ll {\cal E} _{LO}$ ) because we focus on noise analysis. In
that case, the $|{\cal E}_{p,q}|^2$ term can be neglected in
Eq.\ref{DC_term_pq_1}. This means that $D_{p,q}$ gives a good
approximation for the  LO signal.
\begin{equation}\label{DC_term_pq_11}
D_{p,q}  \equiv \sum_{k=1}^{M} I_{k,p,q}   \simeq  M a^2T |{\cal E}_{LO}|^2
\end{equation}
We can get then the signal field  $|{\cal E}_{p,q}|^2$ from Eq.\ref{Sum_heterodyne_2} and Eq.\ref{DC_term_pq_11}:
\begin{equation}\label{Eq_S_ACoverDC}
 \frac{|H_{p,q}|^2}{D_{p,q}} \simeq M a^2 T  |{\cal E}_{p,q}|^2
\end{equation}
In this equation, the ratio ${|H_{p,q}|^2}/{D_{p,q}}$ is
proportional to the number of frames of the sequence ($M=4n$),
This means that ${|H_{p,q}|^2}/{D_{p,q}}$  represents the signal
field $|{\cal E}_{p,q}|^2$ summed over the all frames.

Let us calculate the effect of the shot noise on
${|H_{p,q}|^2}/{D_{p,q}}$. To calculate this effect,  one can make a
Monte Carlo simulation as mentioned above, but a simpler calculation
can be done here. Let us  develop $|H_{p,q}|$ in statistical average
and noise components (as done for $I_{k,p,q}$ in Eq.\ref{EQ_noise on
Ikpq}):
\begin{equation}\label{Eq_nois_on_hpq}
    H_{p,q}= \langle H_{p,q} \rangle  +  h_{p,q}
\end{equation}
with
\begin{equation}\label{Eq_petit_hpq}
    h_{p,q} = \sum_{k=1}^{4n} j^{\;k}  i_{k,p,q}
\end{equation}
Let us calculate $  \langle {|H_{p,q}|^2}/{D_{p,q}} \rangle  $ from
Eq.\ref{Eq_S_ACoverDC}. Since $D_{p,q} \simeq \langle D_{p,q}
\rangle$, we get :
\begin{equation}\label{Eq_S_ACoverDC_222}
\left \langle \frac{|H_{p,q}|^2}{D_{p,q}} \right \rangle \simeq\frac { | \langle H_{p,q} \rangle |^2+ \langle
|h_{p,q}|^2 \rangle + \langle \langle H_{p,q} \rangle h_{p,q}^* \rangle  + \langle \langle H_{p,q}^* \rangle h_{p,q}
\rangle }{ \langle D_{p,q} \rangle}
\end{equation}
In Eq.\ref{Eq_S_ACoverDC_222}  the $\langle \langle H_{p,q} \rangle
h_{p,q}^* \rangle$ term is zero since  $ h_{p,q}^*$ is random while
$\langle H_{p,q} \rangle$ is not. The two terms $\langle \langle
H_{p,q} \rangle h_{p,q}^* \rangle$  and $\langle \langle H_{p,q}^*
\rangle h_{p,q} \rangle$ can be thus removed. On the other hand, we
get for $|h_{p,q}|^2$
\begin{equation}\label{Eq_hpq2}
   |h_{p,q}|^2 = \sum_{k=1}^{4n}   |i_{k,p,q}|^2 +
   \sum_{k=1}^{4n}~~
   \sum_{k'=1,  k' \ne k}^{4n}  j^{~k-k'} i_{k,p,q} i_{k',p,q}
\end{equation}
Since $i_{k,p,q}$ and $ i_{k',p,q}$ are uncorrelated,  the $i_{k,p,q} i_{k',p,q}$ terms cancel in the calculation of
the statistical average of $ |h_{p,q}|^2$. We get then  from Eq.\ref{Eq_Ikpq variance}
\begin{equation}\label{Eq_hpq2_aver}
  \langle  |h_{p,q}|^2 \rangle = \sum_{k=1}^{4n}  \langle |i_{k,p,q}|^2 \rangle
  = \sum_{k=1}^{4n} \langle  I_{k,p,q} \rangle = \langle D_{p,q} \rangle
\end{equation}
Eq.\ref{Eq_S_ACoverDC_222} becomes thus :
\begin{equation}\label{Eq_S_ACoverDC_222_bis}
   \left \langle \frac{|H_{p,q}|^2}{D_{p,q}} \right \rangle  = \frac { | \langle
H_{p,q} \rangle |^2}{ \langle D_{p,q} \rangle} +1
\end{equation}
Equation \ref{Eq_S_ACoverDC_222_bis} means that the average detected
intensity signal $\langle {|H_{p,q}|^2}/{D_{p,q}}\rangle $ is the
sum of the square of the average object field  $\langle
|H_{p,q}|\rangle/({\langle D_{p,q} \rangle}^{1/2}) $ plus one
photo-electron. Without illumination of the object, the average
object field is zero, and the detected signal is 1 photo-electron.
The equation establishes thus that the LO shot noise yields a signal
intensity corresponding exactly to 1 photo-electron (e) per pixel.

The 1 e noise floor, we get here, can be also interpreted as
resulting from the heterodyne detection of the vacuum field
fluctuations~\cite{bachor1998guide}.

\section{Reaching the  Shot Noise in real life holographic experiment.}

In section \ref{section_Theoretical_noise}, we have
shown that the theoretical noise  on the holographic reconstructed
intensity images is 1 photo electron per pixel whatever the number
of recorded frames is. We will now discuss the ability to reach this
limit in real time holographic experiment. Since we consider
implicitly a very weak object beam signal (${\cal E} \ll {\cal E} _{LO}$ ), the noises that must be
considered are the readout noise of the CCD camera, the technical noise from laser amplitude fluctuations on
the LO beam, and the LO beam shot noise, which yields the
theoretical noise limit.
%
%

Consider a typical holographic experiment made with a  PCO Pixelfly 12 bit digital camera. The LO beam power is
adjusted in order to be at half saturation of the digital camera output. Since the camera is 12 bits, and since the
camera ''gain'' is 4.8e/DC,  half saturation corresponds to $2000$~DC on the A/D Converter, i.e. about $10^4$~e on the
each CCD pixel . The LO shot noise, which is about 100~e, is thus much larger than the Pixelfly Read Noise (20~e), Dark
Noise (3~e/sec) and  A/D quantization noise (4.8~e, since 1 DC corresponds to 4.8~e). The noise of the camera can be
neglected, and is not a limiting factor for reaching the noise theoretical limit.

The LO beam  that  reaches the camera is essentially flat field
(i.e. the field intensity $|{\cal E}_{LO}|^2$ is roughly the
same for all the pixels). The LO beam technical noise is thus
highly correlated in all pixels. This is in particular the case
for the noise induced by the fluctuations of the main laser
intensity, or by the vibrations of the mirrors within the LO
beam arm. To illustrate this point, we have recorded a sequence
of $M=4n=4$ frames $I_k$ (with $k=0...3$)  with a LO beam, but
without signal from the object (i.e. without illumination of the
object). We have thus recorded the hologram of the ''vacuum
field''. We have calculated then the complex hologram $H(x,y)$
by Eq.\ref{Sum_heterodyne}, and the reciprocal space hologram
$\tilde H (k_x,k_y)$ by Fourier transform:
\begin{equation}\label{Eq_tilde_H}
   \tilde H (k_x,k_y) = \textrm{FFT}~~H(x,y)
\end{equation}
%
%
%

\begin{figure}[]
\sidecaption
\includegraphics[width=5cm,keepaspectratio=true]{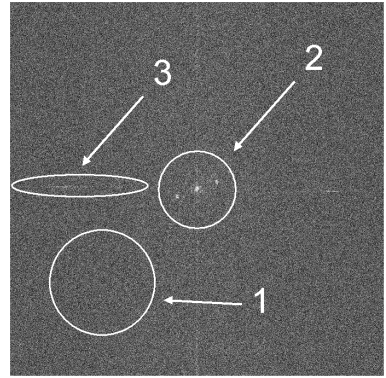}
\caption{ Intensity image of $\tilde{H}(k_x,k_y,0)$  for $M=4n=4$
frames without illumination of the object (no signal field $\cal E$). Three kind of noises
can be identified.  Down left (1) : shot noise; center (2): technical noise of the CCD; left (3) : FFT aliasing. By truncating the
image and keeping only the left down part, the shot noise limit is reached. The image is displayed in arbitrary
logarithm grey scale.
 }\label{fig6_image_shot_noise_zero_signal_exp}
\end{figure}

The reciprocal space holographic  intensity $|\tilde H|^2$  is displayed on
Fig.\ref{fig6_image_shot_noise_zero_signal_exp} in arbitrary logarithm grey scale.  On most of the reciprocal space
(within for example circle~1),  $|\tilde H|^2$ corresponds to a random speckle whose average intensity is uniformly
distributed along $k_x$ and $k_y$. One observes nevertheless  bright points within circle 2, which corresponds to
$(k_x,k_y) \simeq (0,0)$. These points correspond to the technical noise, which is flat field within the CCD plane
$(x,y)$, and which corresponds thus to low spatial frequency components gathered around the center of the $(k_x,k_y)$
reciprocal space. One see also, on the Fig.\ref{fig6_image_shot_noise_zero_signal_exp} image, an horizontal and a
vertical bright line, which corresponds to $k_y\simeq 0$ and $k_x \simeq 0$ (zone 3 on
Fig.\ref{fig6_image_shot_noise_zero_signal_exp}). These parasitic bright lines are related  to Fast Fourier Transform
aliases, that are related to the discontinuity of the signal $I_k$ and $H$ at edge of the calculation grid, in the
$(x,y)$ space.

We have  measured $\langle| {\tilde H}|^2 \rangle$ by  replacing
the statistical average $\langle~\rangle$ by a spatial average
over a region of the conjugate space without technical noise
(i.e. over region~1). This gives a measurement of $\langle|
{\tilde H}|^2 \rangle$, i.e. a measurement of $\langle |  H|^2
\rangle$, since the space average of $| {\tilde H}|^2$ and $| {
H}|^2$ are equal, because of the FFT Parceval theorem. We have
also measured $D$ from the sequence of frames $I_k$ with
$k=0...3$ (see Eq.\ref{DC_term}). Knowing the camera Analog
Digital (A/D) conversion factor (4.8 e/DC), we have calculated
the noise intensity $\langle| {\tilde H}|^2 \rangle/ \langle D
\rangle$ in photo-electron units, and we get, within 10\%, one
photo electron per pixel for the average noise within region~2,
as expected theoretically for the shot noise (see
Eq.\ref{Eq_S_ACoverDC_222}).

To verify that we have truly reached the shot noise limit, we have
performed a control experiment with a camera illuminated by a
tungsten lamp powered by a battery. The lamp  provides here a
clean white light source. The lamp voltage is adjusted to get
half saturation of the camera (about 2000 DC). Like with the
laser experiment described above, we have recorded a sequence of
$M=4n=4$ frames $I_k$ with $k=0...3$, and we have calculated
$H(x,y)$, and $\tilde H(k_x,k_y)$. The image of $|\tilde
H(k_x,k_y)|^2$ we get is very similar to
Fig.\ref{fig6_image_shot_noise_zero_signal_exp}. Moreover, the
average noise intensity in region~2 is exactly  the same as
with a laser (one photo electron per pixel). One has thus :
\begin{equation}\label{Eq_camera_gain}
   {\langle| {\tilde H}|^2 \rangle}/{ \langle D \rangle} =1
\end{equation}
This result is expected since the camera ''gain'' is measured by
assuming that the noise obtained in clean lamp control
experiment is shot noise limited \cite{Newberry1998measuring}.
Assuming Eq.\ref{Eq_camera_gain}, where ${\langle| {\tilde H}|^2
\rangle}/{ \langle D \rangle}$ depends on the camera ''gain'',
is thus equivalent to make a measurement of the  ''gain'' in
e/DC Units. The control experiment made here, redoes the
''gain'' calibration made by the camera manufacturer (i.e.
PCO). We simply get here,  within $10\%$, the same camera gain
($4.8 e/$DC).

%
%

\subsection{Experimental validation with an USAF target.}

\begin{figure}[]
\sidecaption
\includegraphics[width =7.0 cm,keepaspectratio=true]{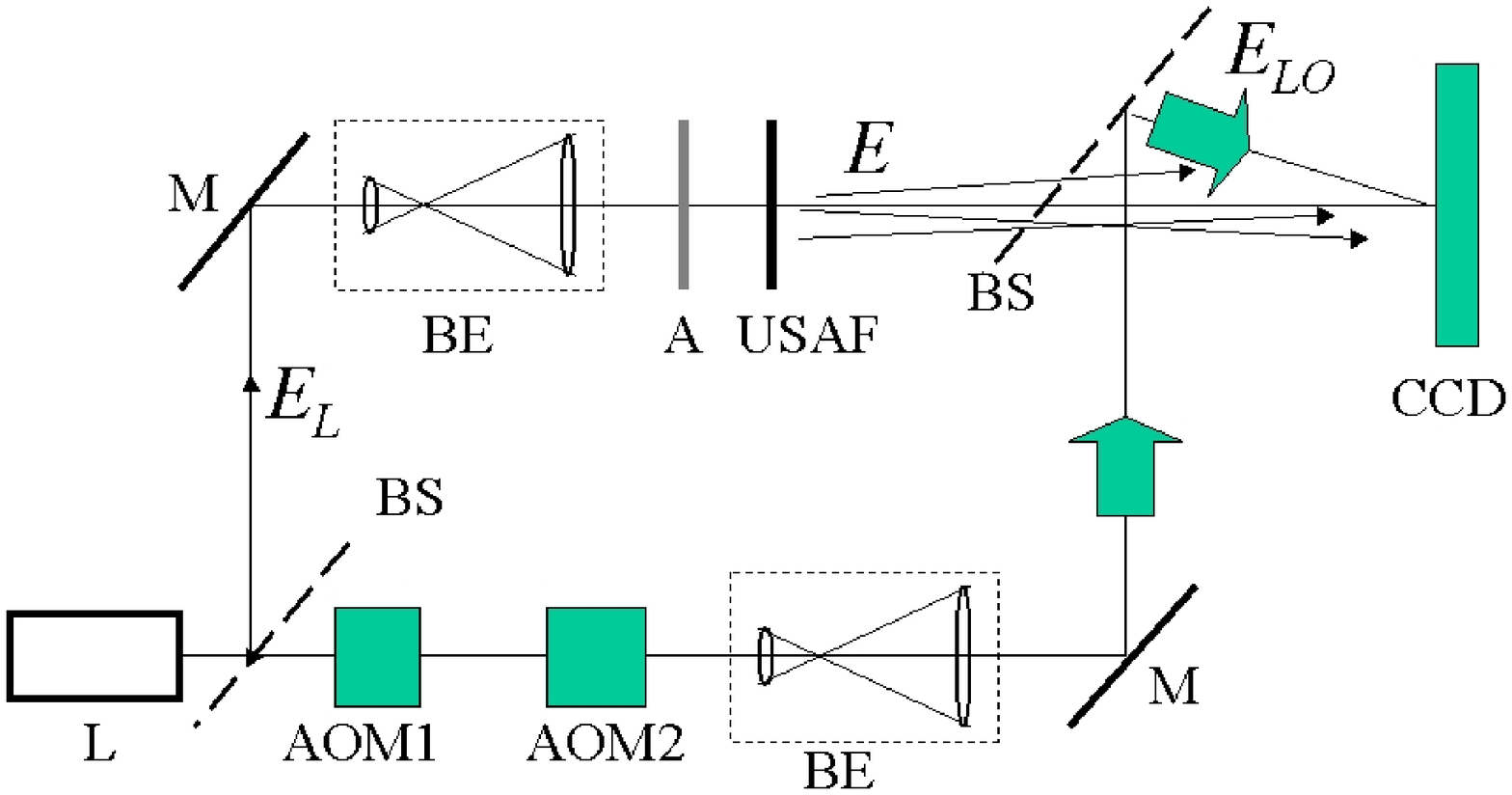}
\caption{ Setup of the test experiment with USAF target. L: main laser; BS: Beam splitter; AOM1 and AOM2: acousto-optic
modulators; BE: beam expander;  M: mirror; A1 and A2: light attenuators. USAF: transmission USAF target that is imaged.
CCD : CCD camera. } \label{fig_setup_usaf}
\end{figure}

We have verified that it is  possible to perform shot noise limited holography in real life, by  recording the hologram of an
USAF target in transmission. The holographic setup is sketched on Fig.\ref{fig_setup_usaf}. We have recorded sequences
of $M=4n=12$ frames, and we have reconstructed the image of the USAF target.

\begin{figure}[]
\sidecaption
\includegraphics[width=10.0cm,keepaspectratio=true]{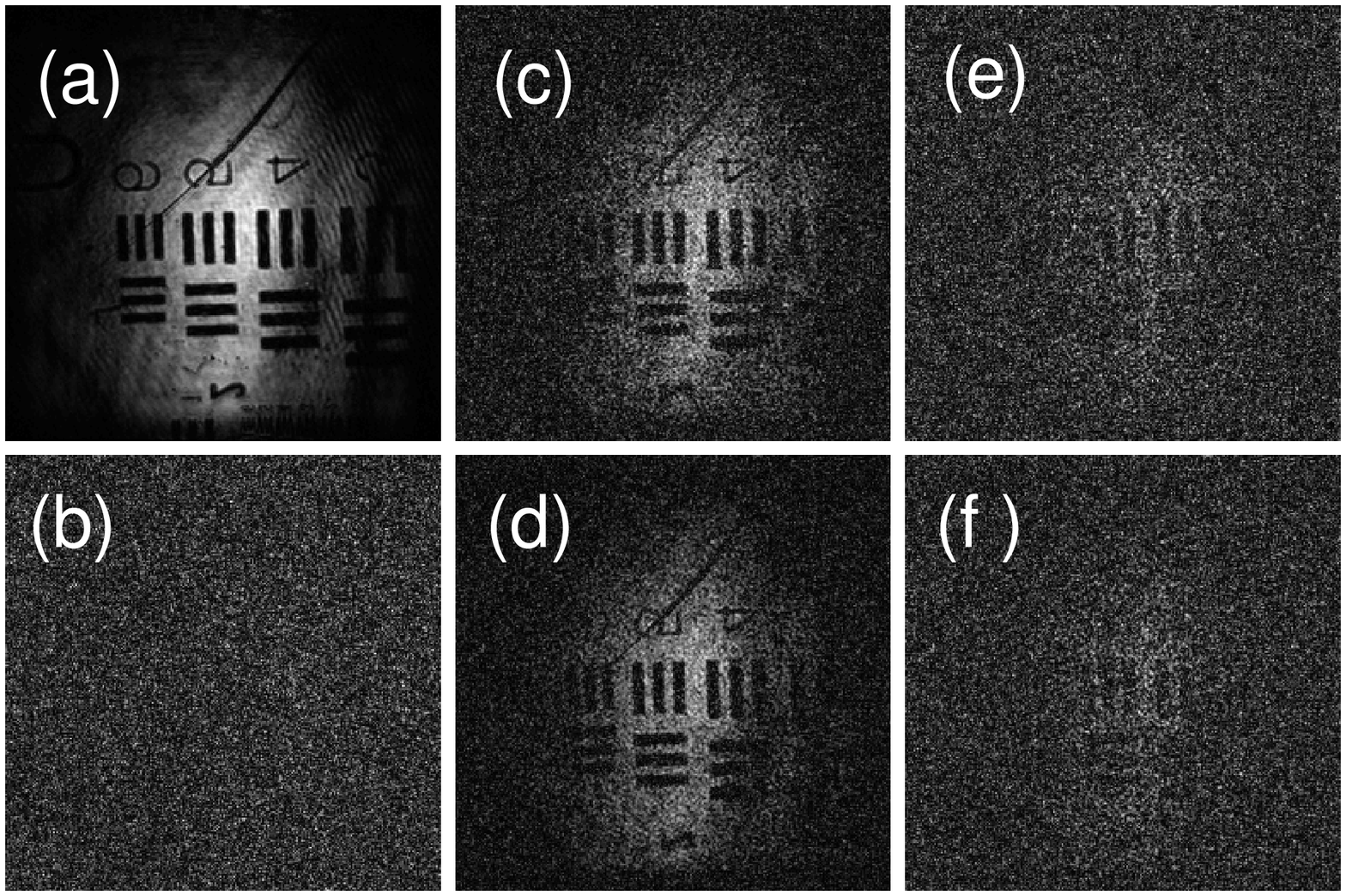}
\caption{(a,c,d): Reconstructions of an USAF target with different level of illumination 700 (a),  1 (c) and 0.15
e/pixel (d). (b): Simulated Shot Noise noise image. (e,f): Simulated reconstructed image obtained by mixing  image (a)
with  weight $X$, and image (b) with weight $1-X$. The weight $X$ is $1/700$ (e), and  $0.15/700$ (f). Images are
displayed in arbitrary logarithmic grey scale.} \label{fig7_usaf_shot}
\end{figure}


Figure \ref{fig7_usaf_shot} shows the holographic reconstructed images of the USAF target.  The intensity of the signal
illumination is adjusted with neutral density filters. In order to filter-off the technical noise, the reconstruction
is done by selecting the order 1 image of the object, within the reciprocal space \cite{Cuche00}. Since the  $400\times
400$ pixels region that is selected is off-axis, the low spatial frequency noisy region, which correspond to the zero
order contributions (region 1 on Fig.\ref{fig6_image_shot_noise_zero_signal_exp}), is filtered-off.

Figure \ref{fig7_usaf_shot} (a,c,d) shows the reconstructed images obtained for different illumination
levels of the USAF target. For each image, we have  measured
%
%
the average number of photo electrons per pixel corresponding to the object beam, within the reciprocal space region
that has been  selected for the reconstruction (i.e. $400\times 400$ pixels). The  images of Fig. \ref{fig7_usaf_shot}
correspond to 700 (a), 1 (c), and 0.15 (d) electron per pixel for the sequence of $M=4n=12$ frames respectively (i.e. $700/12$, $1/12$ and $0.15/12$ e per pixel and  per frame).

Here, the  object beam intensity has been measured by the
following way. We have first calibrated the response of our camera with an attenuated laser whose power is known. We
have then measured  with the camera, at high level of signal, the intensity of the signal beam alone (without LO beam).
We have decreased, to the end, the signal beam intensity by using calibrated attenuators in order to reach the low signal
level of the images of Fig. \ref{fig6_image_shot_noise_zero_signal_exp} (a,c,d). In the case of image (a) with
700e/pix, we also have measured the averaged signal intensity from the data themselves by calculating $|H|^2/D$ (see
Eq.\ref{Eq_S_ACoverDC}). The two measurements gave the same result: 700e per pixel.

On figure \ref{fig7_usaf_shot} (a), with 700e per pixel, the USAF signal is much larger than the  shot noise, and the
Sinal to Noise Ratio (SNR) is large. On figure \ref{fig7_usaf_shot} (c), with 1e per pixel,  the USAF signal roughly
equal to the shot noise, and the SNR is about 1. With 0.15e per pixel, the SNR is low on Fig.\ref{fig7_usaf_shot} (d)
(about 0.15), and the USAF is hardly seen.

It is nevertheless quite difficult to evaluate the SNR of an image. To
perform a more quantitative analysis of the noise within  the images, we have synthesized the noisy images of
Fig.\ref{fig7_usaf_shot} (e,f) by adding noise to the Fig. \ref{fig7_usaf_shot} (a) noiseless image. We have first
synthesized a pure  Noise image, which is displayed on Fig.\ref{fig7_usaf_shot} (b). The Noise image, which corresponds to the image that is expected without signal,  is obtained by the following way.  From one of the
measured frames (for example $I_0$) we have calculated the noise components $i_{k,p,q}$ by Monte Carlo drawing with the
condition:
 \begin{equation}\label{Eq_i_k,p,q_I_0,p,q}
    \langle i_{k,p,q}^2 \rangle = I_{0,p,q}
\end{equation}
This condition corresponds to Eq.\ref{Eq_Ikpq variance} since $\langle I_{k,p,q} \rangle \simeq I_{0,p,q}$. We have synthesized the image sequence $I_k$ in the following manner:
 \begin{equation}\label{Eq_i_k,p,q synthetise}
    I_{k,p,q} = I_{0,p,q} + i_{k,p,q}
\end{equation}
The Shot Noise image of Fig.\ref{fig7_usaf_shot} (b) is reconstructed then from the $I_{k,p,q}$ sequence with
$k=0...12$ since $M=4n=12$.

%
%

We have synthesized noisy images by summing the noiseless image of  Fig.\ref{fig7_usaf_shot} (a) with weight $X$, with
the Noise image of Fig.\ref{fig7_usaf_shot} (b) with weight $(1-X)$.   The image of Fig.  \ref{fig7_usaf_shot} (e)
is obtained with $X=1/700$. Figure  \ref{fig7_usaf_shot} (e) corresponds thus to the same signal, and the same noise than
Figure \ref{fig7_usaf_shot} (c) (i.e. 1e of signal, and 1e of noise respectively). As expected, Fig. \ref{fig7_usaf_shot}
(c) and Fig.  \ref{fig7_usaf_shot} (e) are visually very similar. The image of Fig. \ref{fig7_usaf_shot} is similarly
obtained with $X=0.15/700$. It corresponds to the same Signal and Noise  than Figure \ref{fig7_usaf_shot} (d) (i.e. 0.15e of
signal, and 1e of noise). As expected, Fig. \ref{fig7_usaf_shot} (d) and Fig. \ref{fig7_usaf_shot} (f) are
visually very similar too.

Here we demonstrated  our ability to synthesize a noisy image with a noise that is calculated by Monte Carlo from
Eq.\ref{Eq_i_k,p,q_I_0,p,q} and \ref{Eq_i_k,p,q synthetise}. Moreover, we have verified that the noisy image is
visually equivalent to the image we have obtained in experiments. These results  prove that we are able to
assess quantitatively the noise, and that the noise that is obtained in experiments reaches the
theoretical limit of 1e of noise per pixel for the whole sequence of $M=4n=12 $ frames.
%
%
%

\section{Conclusion}

In this paper we have studied the noise limits in off-axis, heterodyne digital holography. We have shown that because of the heterodyne gain
of the holographic detection, the noise of the CCD camera can be neglected. Moreover by a proper arrangement of the
holographic setup, that combines off-axis  geometry with phase shifting acquisition of holograms by heterodyne holography, it is possible to
reach the theoretical shot noise limit. We have studied theoretically this limit, and we have shown that it corresponds
to 1 photo electron per pixel for the whole sequence of frame that is used to reconstruct the holographic image. This
paradoxical result is related to the heterodyne detection, where the detection bandwidth is inversely proportional to
the measurement time. We have verified all our results experimentally,  and we have shown that is possible to image
objects at very low signal levels. We have also shown that is possible to mimic the very weak illumination levels holograms obtained in experiments by Monte Carlo noise modeling.

%
%
%
%
%

\bibliographystyle{unsrt}

\end{document}